# Evaluation of Moderate Refractive Index Nanoantennas for Enhancing the Photoluminescence Signal of Quantum Dots

Rafael Ramos Uña, Braulio García-Cámara and Ángela I. Barreda*

Department of Electronic Engineering, University Carlos III of Madrid, Avda. de la Universidad, 30, Leganés, 28911 Madrid, Spain

*E-mail: abarreda@ing.uc3m.es

## Abstract

The use of nanostructures to enhance the emission of single-photon sources has withdrawn some attention in last decade due to the development of quantum technologies. In particular, the use of metallic and high refractive index dielectric materials has been proposed. However, the utility of moderate refractive index dielectric nanostructures to achieve more efficient single-photon sources remains unexplored. Here, a systematic comparison of various metallic, high refractive index and moderate refractive index dielectric nanostructures has been performed to optimize the excitation and emission of a CdSe/ZnS single quantum dot at visible spectral region. Several geometries have been evaluated in terms of electric field enhancement and Purcell factor, considering the combination of metallic, high refractive index and moderate refractive index dielectric materials conforming homogeneous and hybrid nanocylinder dimers. Our results demonstrate that moderate refractive index dielectric nanoparticles can enhance the photoluminescence signal of quantum emitters due to their broader electric and magnetic dipolar resonances compared to high refractive index dielectric nanoparticles. However, hybrid combinations of metallic and high refractive index dielectric nanostructures offer the largest intensity enhancement and Purcell factors at the excitation and emission wavelengths of the quantum emitter, respectively. The results of this work may find applications in the development of single-photon sources.

## Introduction

Single-photon sources are a crucial component in the development of photonic integrated circuits for quantum information processing and communication. These sources must generate single photons on demand with high efficiency, purity, and indistinguishability to enable scalable quantum technologies [1-3]. Among other alternatives, semiconductor quantum dots (QDs) are promising candidates for such single-photon sources due to their optical properties, such as narrow emission linewidths, high photon extraction efficiencies, and the potential for electrical injection [4-6]. Additionally, QDs can be engineered to emit at desired wavelengths by tuning their size and composition, making them suitable for various applications in quantum optics and photonics [7, 8]. In particular, QDs emitting in the visible range can be fabricated using various semiconductor materials, such as CdSe, CdTe, InP [9] or recently perovskites [10]. These QDs can be synthesized using colloidal chemistry techniques, which allow for precise control over their size, shape, and composition, even they can be embedded in a wide range of host materials, such as polymers or silica, to improve their stability and ease of integration into photonic devices [11, 12].

Despite the promising properties of visible-range QDs for single-photon sources, the stringent requirements of practical applications still require remarkable efforts to improve the efficiency and purity [13]. In this sense, one of the main challenges is to enhance the spontaneous emission rate of the QD, which determines the photon generation efficiency and timing jitter. This can be achieved by exploiting the Purcell effect, which occurs when the QD is coupled to a resonant nanostructure that modifies the local density of optical states (LDOS) experienced by the emitter. In this regard, there are a huge number of proposals in the state of the art considering different materials and geometries [14-16]. In particular, plasmonic nanostructures are well known by their capacity to confine light into deep subwavelength volumes, leading to strong field enhancements and large Purcell factors. This enhancement can increase the spontaneous emission rate of the QD, improving the photon generation efficiency and reducing the time response [17, 18]. However, plasmonic structures also introduce additional losses, which can degrade the quantum efficiency and indistinguishability of the emitted photons. An alternative approach is to use high refractive index (HRI) dielectric nanostructures, which can support Mie resonances and exhibit low ohmic losses in the visible range compared to plasmonic structures [19, 20]. These resonances can also lead to strong field enhancements and large Purcell factors, while maintaining high quantum efficiencies. Furthermore, the resonant frequencies of dielectric nanostructures can be tuned by varying their size, shape and material composition, allowing for optimal coupling with the QD emission. Additionally, dielectric nanostructures can be fabricated using well-established semiconductor processing techniques, making them compatible with existing photonic integrated circuit technologies [21, 22]. However, the field enhancement in dielectric nanostructures is typically lower than in plasmonic structures due to the weaker light confinement. To overcome this limitation, researchers have explored the use of hybrid plasmonic-dielectric nanostructures, which combine the advantages of both materials [23-26]. These structures can exhibit remarkable field enhancements while maintaining low losses, making them promising candidates for enhancing the performance of QD-based single-photon sources.

Recently, a new alternative has arisen with the use moderate refractive index (MRI) dielectric nanostructures [27]. MRI (1.7 < n < 3) dielectric nanoparticles (NPs), typically made from materials like silicon nitride ($Si_3N_4$) or titanium dioxide ($TiO_2$), offer several advantages over their HRI counterparts, such as being chemically synthesized and flexibly assembled [28]. The resonances of MRI are located at shorter wavelengths in comparison with those of HRI dielectric nanospheres with same size and shape, because their spectral evolution, mainly the magnetic dipolar resonance, follows the formula $\lambda \approx n \cdot d$, being $\lambda$ the resonant wavelength, $n$ the refractive index, and $d$ the diameter of the nanosphere [29]. Moreover, previous studies have already demonstrated that MRI dielectric NPs show broad Mie resonances in the visible spectral range. In [30] it was evidenced that nanospheres with refractive index 2.2, i.e. synthetized hydrogenated amorphous silicon (a-Si:H), show broad resonant conditions due to the overlapping of the dipolar electric and magnetic Mie modes, leading to emission enhancement of several excitons of 2D materials located at the bottom of the MRI NP. Here, we propose a comparative study of different nanostructures to optimize both the excitation and emission of single QDs. In particular, the proposed structures consist of homogeneous and hybrid dimers made of metallic, MRI and HRI dielectric materials. The results of this work can have applications in the development of more efficient single-photon sources.

## Methodology

The presented results were obtained by means of the commercial software COMSOL Multiphysics, which employs the Finite Element Method (FEM) to solve the Maxwell equations by stablishing the boundary conditions of the system. The considered geometry for the nanoantenna consists of a dimer of nanocylinders. As it can be observed in Fig. 1, two different cases have been analyzed. On the one hand, homogeneous dimers of metallic, MRI or HRI dielectric NPs. On the other hand, hybrid nanostructures combining metallic and dielectric NPs. Silicon and gold were considered as the HRI and metallic materials, respectively, being their theoretical optical properties obtained from Palik [31], while MRI had a constant refractive index of 2.2 in all the analyzed spectral range. Glass was considered as the material for the substrate, with a refractive index of 1.5.

The dimer configuration was set to enhance the excitation and emission of a quantum emitter, corresponding to a QD located in the middle of the gap between the cylinders over the substrate, as this may be the most suitable location for the QD. The gap was set to be 20 nm wide due to fabrication limitations, and the QD was established to have a radius of 5 nm according to the sizes provided in the literature [32-35]. Specifically, a CdSe/ZnS QD is considered, whose excitation and emission wavelengths are centered at 570 and 650 nm, respectively [35, 36].

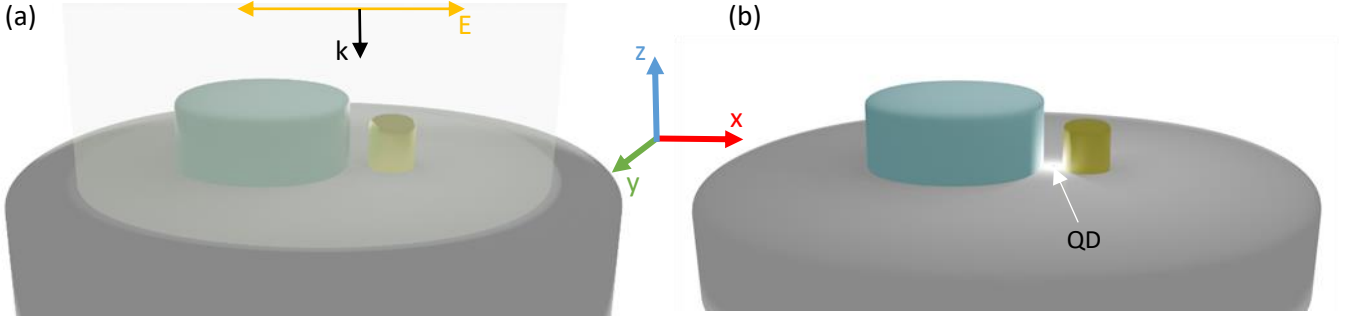

**Figure 1.** (a) Illumination of the dimer by a plane wave propagating along the negative direction of the *z*-axis, i.e., from the air to the substrate, and linearly polarized parallel to the dimer orientation (*x*-axis). (b) Dimer is excited by an electric dipole oscillating in the *x*-direction, which represents the QD, located in the middle of the gap and 5 nm above the substrate to take into account the QD size.

The most suitable radii and height for the cylinders of the dimers have been selected with the main objective of obtaining plasmonic or Mie resonances at the excitation and emission wavelengths. To have an idea of the approximate sizes of the NPs in the dimer, firstly, we have selected the sizes of the single metallic and dielectric NPs showing the dipolar resonances in the spectral region of interest. In particular, a multipolar decomposition of every selected nanocylinder was performed in order to obtain the modes contributions to the scattering efficiency. The different contributions were obtained by integrating the displacement currents induced within the NPs [37]. It is important to mention that the multipolar decompositions have been performed in a homogeneous media, without a substrate. Once selected the sizes that best adapt for every material, dimers were studied in terms of excitation and emission enhancement. The excitation enhancement is given by the square module of the electric field enhancement $|E|^2/|E_0|^2$ at the excitation wavelength; here $|E|$ is the electric field amplitude averaged in the volume of the QD, and $|E_0|$ corresponds to the electric field amplitude of the incident plane wave averaged in the same volume. The emission enhancement is provided by the Purcell factor ($F$).

As both excitation and emission enhancements were evaluated, two sets of simulations have been carried out to analyze the performance of the different proposed geometries (see Fig. 2). For the calculation of the scattering efficiency spectra and the electric field enhancement at the QD position due to the nanoantenna, the NPs were illuminated by a plane wave linearly polarized along the *x*-axis, i.e. the axis that joins both cylinders of the dimer configuration, and propagating along the negative direction of the *z*-axis, i.e. the disk axis. While to assess the Purcell factor and radiation efficiency, it was required to incorporate the QD into the simulations. The QD is represented by an electric point dipole source, with a dipole moment oriented along the *x*-axis, placed in the gap between the cylinders and 5 nm above the substrate to take into account the QD size, as described before. According to the Purcell effect, the spontaneous emission rate of a quantum emitter can be modified due to its interaction with a resonant optical cavity [38]. In particular, the total decay rate of a quantum emitter inside a cavity is driven by the Purcell factor $F = 3/(4\pi^2) \cdot Q/V$, where $V$ is the volume normalized by the wavelength ($\lambda$) over the local mode refractive index ($n$) cubed $(\lambda/n)^3$ and $Q$ is the resonator quality factor [38-42]. To assess the Purcell factor, which is related to LDOS, the following expression was considered:

$$F = \frac{P_r + P_{nr}}{P_r^0 + P_{nr}^0} \quad (1)$$

where $P_r$ is the radiative power, $P_{nr}$ the non-radiative power, and $P_r^0$ and $P_{nr}^0$ their intrinsic counterparts. This relation describes the total power emitted by the dipole along with the nanoantenna over the power emitted by the dipole in air.

The radiation efficiency is calculated as the radiative power over the total emitted power (radiative and non-radiative):

$$\eta = \frac{P_r}{P_r + P_{nr}} \quad (2)$$

For performing the simulations in COMSOL, a sphere of radius 800 nm was surrounding the geometry, in order to calculate the scattering cross section and the radiative power, by integrating the Poynting vector at its surface. The non-radiative power was obtained by integrating the ohmic losses in the volume of the NPs. A sphere of radius 1200 nm was divided into two halves, the upper one was made of air, and the lower one was made of the substrate refractive index. This sphere was surrounded by a perfectly matched layer (PML) of thickness $\lambda/4$, being $\lambda$ the wavelength of the incident radiation.

As previously mentioned, for the scattering efficiency and electric field enhancement simulations, a plane wave was chosen as the illumination source. However, for the Purcell factor calculations, a dipole source was used. To get the convergency of the results, when the dipolar source is included, a sphere of radius 5 nm surrounding the dipole was created. Its mesh size corresponds to 0.6 nm as maximum element size and 0.03 nm as minimum.

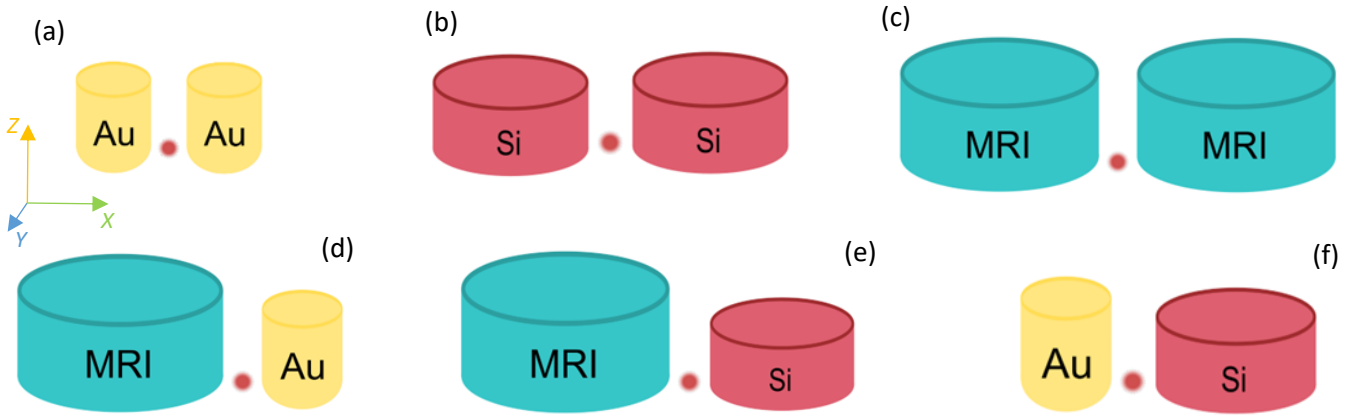

**Figure 2**. Scheme of all different dimer configuration. From (a) to (c) non-hybrid (homogeneous) dimers are represented: (a) two gold NPs, (b) two silicon NPs and (c) two MRI NPs. And from (d) to (f) hybrid dimers combining all three previous materials. The red dot between the cylinders represents the CdSe/ZnS QD.

## Results

A comparative study between metallic, HRI and MRI dielectric materials for a nanocylindrical dimer has been performed in order to increase the Purcell factor, efficiency and electric field enhancement. Larger enhancements are met when the gap between the cylinders is reduced, however, the gap considered was 20 nm, as it can be fabricated by standard electron-beam lithography. Also, sizes for the nanocylinders were limited in the range of 20-200 nm in height and 50-300 nm in radius, due not only to manufacturing limitations but also to the necessity of staying in a subwavelength range.

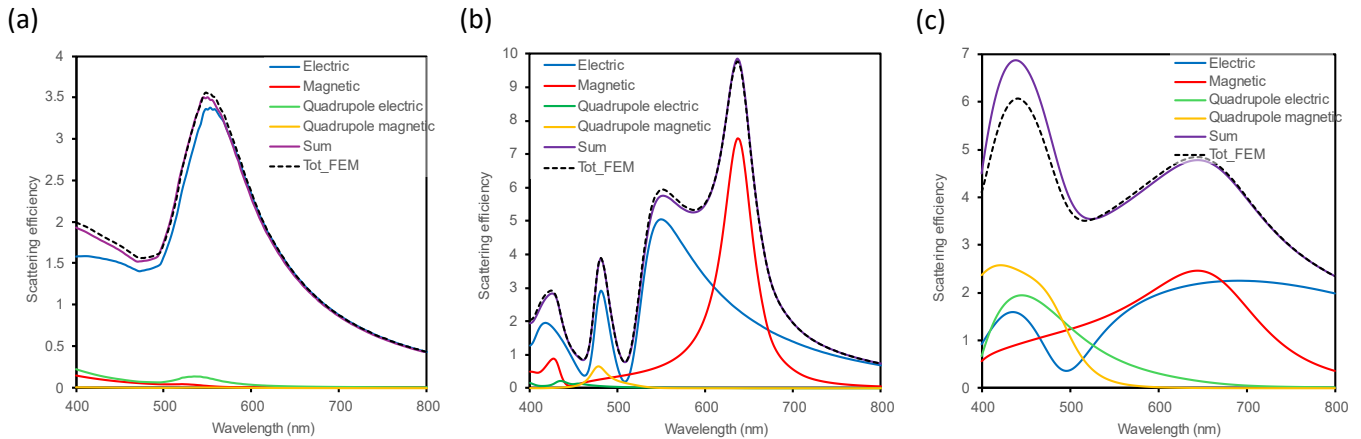

**Figure 3.** Scattering efficiency and multipolar decomposition for metallic and dielectric isolated cylinders in air. The nanoparticles have been illuminated by a plane wave propagating along the negative direction of the $z$-axis (parallel to the disk axis) and linearly polarized along the $x$-axis (perpendicular to the disk axis). (a) Gold nanocylinder of $R = 50$ nm and $H = 150$ nm, (b) Silicon nanocylinder of $R = 120$ nm and $H = 80$ nm and (c) MRI nanocylinder of $R = 150$ nm and $H = 200$ nm. Black sliced line: scattering efficiency obtained with COMSOL. Purple solid line: scattering efficiency obtained as the sum of the dipolar electric and magnetic, and quadrupolar electric and magnetic contributions. Blue and red solid lines: dipolar electric and magnetic contributions, respectively. Green and yellow solid lines: quadrupolar electric and magnetic contributions, respectively.

*3.1 Single nanocylinders*

Several sizes for the single metallic, HRI and MRI NP have been considered in order to attain the appropriate radius and height of the cylinder that provides the dipolar resonances at the excitation and emission wavelengths of the QD. However, for simplicity, only the results for the chosen dimensions are represented. In Fig. 3 the multipolar decomposition is shown for a metallic cylinder of radius $R = 50$ nm and height $H = 150$ nm, a HRI dielectric one of $R = 80$ nm and $H = 120$ nm, and a MRI dielectric cylinder of $R = 150$ nm and $H = 200$ nm.

As it is observed in Fig. 3, while the multipolar decomposition for the gold NP only shows electric response, for the dielectric ones both magnetic and electric dipolar resonances are attained. Focusing on the regions of interest, i.e. in the surroundings of 570 nm and 650 nm, the dipolar response dominates. For the gold NP, it is observed an electric dipolar resonance at the excitation wavelength. In the case of the HRI dielectric NP, at the excitation and emission wavelengths, the main contributions to the scattering efficiency come from the dipolar electric and magnetic resonances. Differently, for the MRI dielectric NP, dipolar electric and magnetic resonances are broader than for the HRI dielectric one, achieving almost the same contribution of both resonances to the total scattering in the spectral range between 550 and 700 nm.

*3.2 Homogeneous dimers*

The sizes of the single NPs were optimized to achieve the dipolar electric and magnetic response within the wavelength range of interest, as explained. Therefore, the dimensions were kept constant when forming the dimers. Even known the presence of a substrate along with the interaction of both NPs generate a red-shift of the resonances, the scattering of the dimers adjusts to the desired wavelength range.

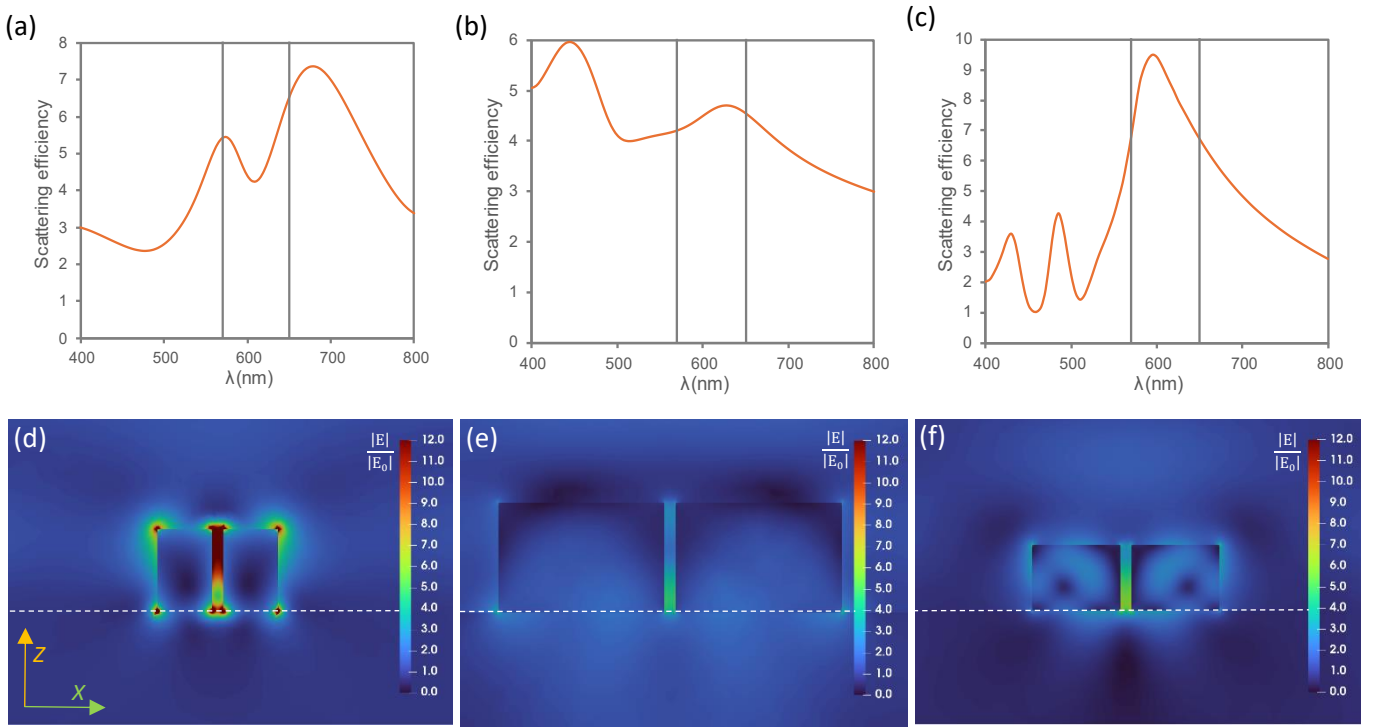

**Figure 4.** (a) to (c) Scattering efficiency for homogeneous dimers: (a) Au-Au, (b) MRI-MRI and (c) Si-Si dimers. The gray vertical lines correspond to the excitation and emission wavelengths. (d) to (f) Near field maps (Z-X plane) at the excitation wavelength for the homogeneous dimers: (d) Au-Au, (e) MRI-MRI and (f) Si-Si. A hot-spot is observed in between the cylinders, specified as the reddish/greenish area. The dashed white line represents the position of the substrate.

3.2.1 Electric field enhancement and scattering efficiency spectra

In line with the previous section, scattering efficiency spectra was evaluated for the homogeneous dimers. In Fig. 4 a-c, it is observed that the scattering efficiency spectra for the dimers are broader than for the isolated NPs due to the interaction effects between both components of the dimer and the substrate presence, which was considered for the dimer analysis. Furthermore, the resonances are attained in between the excitation and emission wavelengths, leading to high scattering efficiencies at $\lambda = 570$ nm and $\lambda = 650$ nm. Comparing Fig. 4b and 4c, it is evidenced that the MRI dimer show broader resonances, but lower scattering efficiencies.

Focusing on the electric field enhancement, in Fig. 4 d-f the near-field maps are represented in the scattering plane (Z-X) at the excitation wavelength, for the corresponding dimers. The highest excitation enhancement is attained for the gold dimer. In spite of that, hot-spots were also observed in the gap between the HRI and MRI dielectric NPs, being this the main reason of analyzing dimers instead of single NPs, where the strongest enhancement is achieved inside the NPs. Specifically, from the near-field maps, it is shown that a hot-spot is localized in between the cylinders encompassing the volume of the QD. The results for the average field enhancement over the volume of the QD are shown in Table 1, where the largest enhancement was achieved for the gold dimer, followed by the silicon and then the MRI one.

**Table 1.** Purcell factor ($F$), radiation efficiency ($\eta$), averaged electric field enhancement at the position of the QD and product of the excitation and emission enhancement for the considered homogeneous and hybrid dimers. For a more direct comparison the same size for each material is used in every geometry as in Fig. 3.

|  |  |  | NP 1 |  | NP 2 |  |  |  |  |  |
| --- | --- | --- | --- | --- | --- | --- | --- | --- | --- | --- |
|  |  | Materials | $R_1$ [nm] | $H_1$ [nm] | $R_2$ [nm] | $H_2$ [nm] | **F** | $\eta$ | **$|E|/|E_0|$** | **$|E|/|E_0|^2 \cdot F$** |
| Non-hybrid dimers | #1 | Au-Au | 50 | 150 | 50 | 150 | **285.64** | 0.66 | **8.612** | **32953** |
|  | #2 | MRI-MRI | 150 | 200 | 150 | 200 | **7.63** | 1 | **3.18** | **120** |
|  | #3 | Si-Si | 80 | 120 | 80 | 120 | **31.80** | 0.96 | **5.39** | **1435** |
| Hybrid dimers | #4 | MRI-Au | 150 | 200 | 50 | 150 | **109.47** | 0.66 | **4.64** | **2672** |
|  | #5 | MRI-Si | 150 | 200 | 80 | 120 | **18.31** | 0.97 | **4.06** | **470** |
|  | #6 | Si-Au | 80 | 120 | 50 | 150 | **138.39** | 0.72 | **7.44** | **11916** |

3.2.2 Purcell factor and efficiency

Concerning the study of emission enhancement of the QD, an electric dipole located in the gap between the cylinders of the dimer was used as the light source, as it is described in the methodology section. The presence of an antenna usually boosts the LDOS at the emission wavelength, decreasing the radiative lifetime. In particular, a high Purcell factor is desired to accelerate the emission process. Moreover, large radiation efficiency is also requested in order to attain efficient single quantum emitters. In Table 1, it is summarized the Purcell factor and the efficiency for the different studied dimers.

The Purcell factor, calculated as described in Eq. 1, demonstrates that the emission of the QD is more significantly influenced by the presence of the gold dimer than by the dielectric NPs. Specifically, the Purcell factor is almost 9 times larger for gold than for silicon, and more than 37 times larger than for the MRI dimer. Regarding the efficiency, the highest values were obtained for the dielectric materials, due to the low losses in comparison with the metallic ones that suffer the Joule effect. It is worth noticing that the MRI NPs are considered to have a negligible imaginary part of the refractive index ($k = 0$), being this the reason for that 100% of radiative emission.

The product of the excitation (square of the electric field enhancement) and emission enhancement (Purcell factor) is shown in the last column of Table 1. Based on the results, it is concluded that the performance of MRI nanoantennas is not as effective as that of metallic or HRI nanoantennas for designing efficient quantum emitters.

*3.3 Hybrid dimers*

Hybrid nanostructures have been analyzed due to their ability to combine the best properties from dielectric and metallic NPs, strong confinements of electromagnetic energy in the surroundings of metallic NPs and low losses of dielectric ones.

3.3.1 Electric field enhancement and scattering efficiency spectra

The scattering efficiency for each material combination is represented in Fig. 5 a-c, where again the combinations involving a MRI dielectric nanocylinder show broad Mie resonances at the desired wavelength range, but higher scattering efficiencies are reached when employing the HRI dielectric NPs. Near field maps at excitation wavelength (570 nm) are plotted in Fig. 5 d-f. The highest hot-spots are observed in the corners of the NPs due to the lightning rod effect. However, still hot-spots are achievable at the QD position. In Table 1, it is shown the results for the average field enhancement over the volume of the QD, being the largest enhancement attained for the gold-silicon combination.

### 3.2.2 Purcell factor and efficiency

The results regarding Purcell factor and efficiency for these hybrid combinations are shown in Table 1. It is firstly observed that a combination of MRI with either silicon or gold increases the product of the excitation and emission enhancement with respect to the MRI dimer, but it is still lower than the HRI-gold dimer combination.

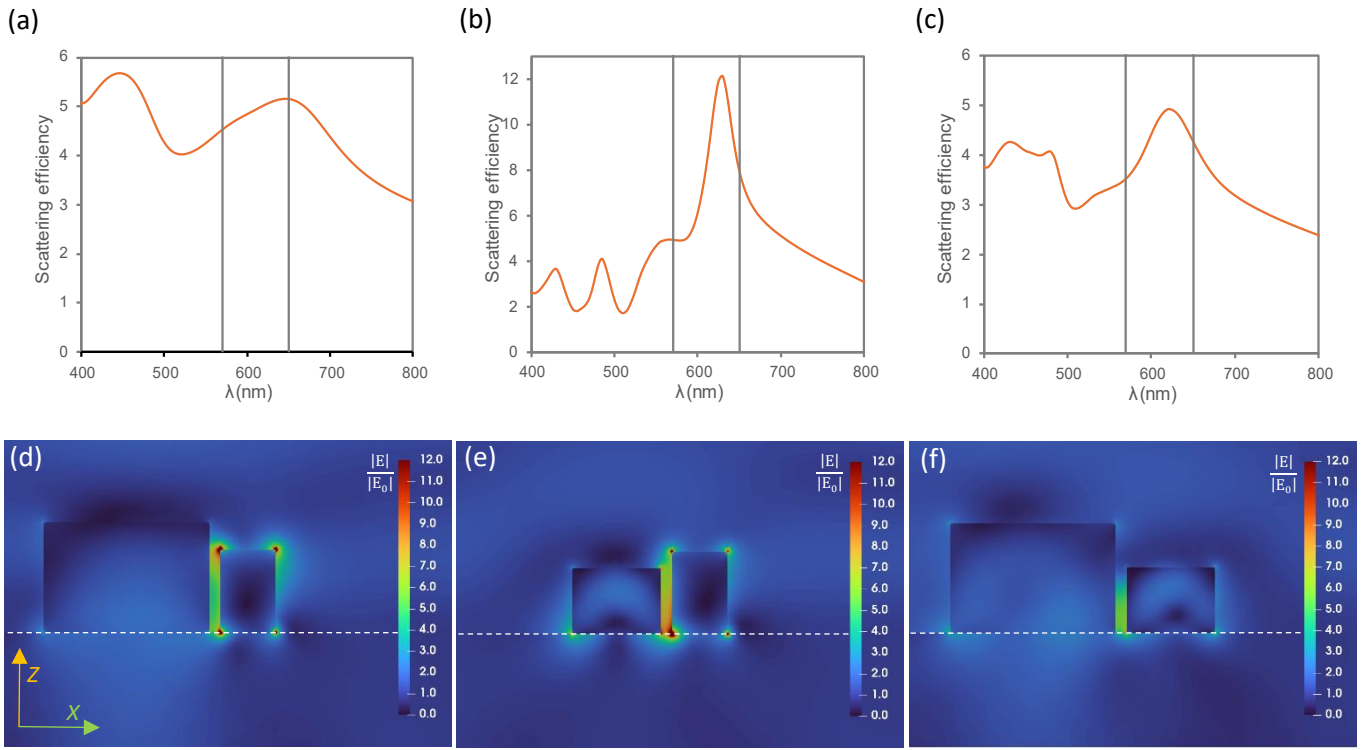

**Figure 5.** (a) to (c) Scattering efficiency for hybrid dimers: (a) MRI-Au, (b) MRI-Si and (c) Au-Si dimers. The gray vertical lines correspond to the excitation and emission wavelengths. (d) to (f) Near field maps (Z-X plane) at the excitation wavelength for the hybrid dimers: (d) MRI-Au, (e) MRI-Si and (f) Au-Si. The dashed white line represents the position of the substrate.

## Conclusions

Studies have demonstrated that metallic and high refractive index dielectric nanoparticles are effective in enhancing the emission of quantum emitters, with important applications in quantum communications and quantum information. However, the case of moderate refractive index dielectric materials remains unexplored for this objective. In this work, we have carried out a systematic comparison of homogenous and hybrid dimers of metallic, high refractive index and moderate refractive index dielectric nanoparticles for enhancing the excitation and emission of a quantum dot located in the gap between the nanoparticles.

Firstly, we have optimized the size of the single metallic, high refractive index and moderate refractive index dielectric nanoparticles to attain dipolar resonances at the wavelength range of interest. Secondly, homogeneous and hybrid dimers have been analyzed for a fixed gap of 20 nm to fulfil fabrication constraints. It has been observed that although an overlapping of electric and magnetic Mie resonances appears for the moderate refractive index dielectric materials, leading to broader resonances than those attained for metallic or high refractive index dielectric nanoparticles, the excitation and emission enhancement is lower than for those other materials. In particular, the largest electric field enhancements at the position of the quantum dot and Purcell factors are attained for the gold dimer. However, low efficiencies are reported due to the ohmic losses. Silicon dimers provide lower excitation enhancement and Purcell factors than the gold dimers, nevertheless, this low value is compensated by the larger efficiency due to the low losses. Regarding the hybrid nanostructures, the combination of silicon-gold provides better results than those involving moderate refractive index dielectric nanoparticles. This leads to conclude that moderate refractive index dielectric nanoparticles do not perform as well as metallic and high refractive index dielectric materials in increasing the efficiency of single quantum emitters.

However, they can still be interesting structures for enhancing the emission of multiple excitons in 2D materials due to their broadband electromagnetic response.

**Acknowledgements**

A.B. thanks MICINN for the Ramón y Cajal Fellowship (grant No. RYC2021-030880-I). R.R thanks also the Spanish Research Agency (AEI) for his predoctoral grant (PID2022-137857NA-I00).

**Funding**

This work was funded by the Spanish Research Agency (AEI) through the project No. PID2022-137857NA-100.